\pdfoutput=1
\documentclass[cits]{PoS}

\usepackage{amsmath}
\usepackage{slashed}
\usepackage{url}

\newcommand{\Li}[0]{\mathrm{Li}}
\newcommand{\tbet}{{\tilde{\beta}}}

\title{The new PV prescription for IR singularities of NLO splitting functions} 

\ShortTitle{The new PV prescription for IR singularities of NLO splitting
functions} 

\author{\speaker{M.\ Skrzypek}$^a$, O.\ Gituliar$^{b}$, S.\ Jadach$^a$,
A.~Kusina$^c$ \\
\llap{$^a$} Institute of Nuclear Physics, Polish Academy of Sciences,\\
              ul.\ Radzikowskiego 152, 31-342 Cracow, Poland\\
\llap{$^b$} DESY, Platanenallee 6, D-15738 Zeuthen, Germany\\
\llap{$^c$} Southern Methodist University, Dallas, TX 75275, USA\\
E-mail: \email{Maciej.Skrzypek@ifj.edu.pl}, \email{Oleksandr.Gituliar@desy.de},
\email{Stanislaw.Jadach@ifj.edu.pl}, \email{akusina@smu.edu}
}
\abstract{In this note we outline the Monte Carlo project {\tt KrkMC}. The goal
of
this project is to construct a QCD Parton Shower accurate to NLO level in both
coefficient function and splitting function (shower) parts. We discuss in
detail one of its aspects --- the evolution kernels. The kernels had to be
recalculated in a new regularisation scheme, called NPV. In this
scheme all
the singularities in the plus component of the integration momenta are
regularised by means of principal value
prescription. This is in contrast to the standard approach, in which only the
spurious axial singularities are regularised by principal value. As a result,
the triple
poles in the dimensional regularisation parameter $\epsilon$ are replaced by a
combination
of $\epsilon$-poles and logarithms of geometrical cut-off
$\delta$. The resulting exclusive parton densities are more suitable
for stochastic applications in four dimensions. Simultaneously, at the
inclusive level, the standard and new prescriptions give the same results
provided appropriate real and virtual contributions are added.
\\
  \begin{flushright}
    DESY 14-115 \\
    IFJPAN-IV-2014-11 \\
    LPN14-083 \\
    SMU-HEP-14-04 
  \end{flushright}
}

\FullConference{Loops and Legs in Quantum Field Theory - LL 2014,\\ 
          27 April - 2 May 2014 \\ 
          Weimar, Germany}

\begin{document} 

\section{Introduction}
Recently the LHC collaborations managed to demonstrate
that Higgs boson decays to fermions are as predicted by the
Standard Model \cite{Chatrchyan:2014vua}.
These results mark the onset of the era of "precision Higgs physics"
\cite{CERNpress_precisionLHC}, already before the beginning of LHC Run 2,
scheduled for early 2015. Such spectacular experimental results call for
adequate theoretical tools for stochastic simulations of $pp$ collisions,
ultimately to NNLO accuracy. The development of Monte Carlo parton
showers started in the mid 1980-ies with the (improved) LO shower and hard
processes
\cite{Sjostrand:1985xi,Webber:1984if}. It took almost twenty years, until the
mid
2000-s, for the next step: hard processes upgraded to NLO precision, while
retaining the LO shower \cite{Nason:2004rx,Frixione:2007vw,Frixione:2002ik}, and
this is the present day state-of-the-art solution. For comparison; the
analytical inclusive results reached the NNLO precision ten years ago
\cite{Moch:2004pa,Vogt:2004mw,Almasy:2011eq,Mitov:2006ic}.

\section{The {\tt KrkMC} project}
There are a number of promising ideas on how to improve the precision of parton
showers
\cite{Tsuno:2006cu,Hoeche:2012yf,Alioli:2012fc,
Hamilton:2012np, Nagy:2014mqa,Nagy:2014nqa}.
We will focus here on the {\tt
KrkMC} project, developed in Krak\'ow
\cite{Jadach:2011cr,Skrzypek:2011zw,Jadach:2012vs,Jadach:2013dfd,Jadach:2009gm,
Jadach:2010ew}. 
Its goal is to propose a new, complete scheme of NLO shower and NLO hard
processes. Let us outline briefly its major features.
It is based on collinear factorization
\cite{Ellis:1978ty,Ellis:1978sf,Curci:1980uw,Collins:1984kg}, which provides a
solid field-theoretical basis. Of course, the collinear factorization is
"inclusive"; the transverse degrees of freedom are integrated out. Therefore the
new scheme requires:
\begin{itemize}
\item
Reformulation of collinear factorization theorem in a fully exclusive way.
\item
Recalculation of the evolution kernels:
in an exclusive way,
in four dimensions,
with well defined relation to $\overline{\text{MS}}$ kernels.
\item
Construction of kinematical mappings to the true phase space without any gaps
or overlaps.
\item
Formulation of reweighting procedure, reasonably convergent, with positive
weights.
\end{itemize}
As an illustration, let us present a general formula used as a basis of the
new NLO scheme; both for the case of upgrading hard matrix elements as well as
showers. We show the formulae in a schematic way and we refer to
\cite{Jadach:2011cr,Jadach:2011kc} for details. In both cases, the starting
point is the
LO Monte Carlo, that we represent symbolically as (only one hemisphere,
$B$ -- backward, is shown)
\begin{equation}
\sigma(s) =
\int dx d\sigma_0({xs},\hat\theta)
e^{-S_{_{ISR}}}
\bigg[
\delta_{x=1} +
\sum\limits_{n=1}^\infty\;
\bigg( \prod\limits_{i=1}^n\; 
    \int\limits_{Q>a_i>a_{i-1}}\!\!\!\!\!\!  d^3\eta_i\;
    \rho^{(1)}_{1B}(k_i)
\bigg)
\delta_{x=\prod_{j=1}^n x_j}
\bigg]
\end{equation}
where the $\rho$ function is the LO distribution.
The inclusion of the NLO correction to the hard process is done by means of a
simple weight 
\begin{equation}
W^{NLO}_{MC}=
1+\Delta_{S+V}
+\sum_{j\in B} 
 \frac{\tbet_1(\hat{s},\hat{p}_F,\hat{p}_B;a_j, z_{Bj})}%
      {\bar{P}(z_{Bj})\;d\sigma_0(\hat{s},\hat\theta)/d\Omega}
+ \{B\leftrightarrow F\}
\end{equation}
where $\bar{P}(z)\equiv \frac{1}{2}(1+z^2)$ and
\begin{eqnarray}
&
\tbet_1(\hat{p}_F,\hat{p}_B;q_1,q_2,k)=
  \frac{(1-\beta)^2}{2}
  \frac{d\sigma_{0}}{d\Omega_q}(\hat{s},\theta_{B2})
-\theta_{\alpha<\beta}
 \frac{1+(1-\alpha-\beta)^2}{2}
 \frac{d\sigma_{0}}{d\Omega_q}(\hat{s},\hat\theta)
+ \{B\leftrightarrow F\}
\end{eqnarray}
is the {IR- and collinear-finite} {real} emission part and
$\Delta_{V+S}$ is a constant
{\em virtual+soft} correction.
%
The inclusion of the NLO $C_F^2$-type corrections in the shower is also done by
means of a simple weight
\begin{eqnarray}
&\bar W^{C_F^2}_{MC}=
1+\sum\limits_{p=1}^{n}
{\beta_0^{(1)}(z_p)}
+ \sum\limits_{p=2}^{n} \sum\limits_{j=1}^{p-1}{W(\tilde{k}_p, \tilde{k}_j)}.
\end{eqnarray}
The double sum represents summation over the whole cascade of the real-real
corrections $W$ that depend on two four-momenta, see~\cite{Jadach:2011kc} for details.
The single sum of $\beta_0$-functions corresponds to real-virtual contributions.
The rest of this note will be devoted to calculation of these real-virtual
contributions to the $P_{qq}$ kernel, see
also \cite{Gituliar:2014wha,Gituliar:2014mua,Gituliar:2013eka,Gituliar:2013rta}.

\section{PV prescription}
In the collinear factorization the use of the axial gauge is instrumental.
It leads to a transparent physical picture of the evolution and allows
for its interpretation in the language of the parton shower. 
On the other hand though, the axial gauge introduces 
spurious (unphysical) singularities at the intermediate steps of the
calculations. Namely, the gluon propagator has the form
\begin{equation}
\nonumber
\frac{1}{l^2}\Bigl(g^{\mu\nu} -\frac{l^\mu n^\nu + n^\mu l^\nu}{nl}
\Bigr)
\end{equation}
where  $n$ is the light-like gauge vector and the denominator $1/nl$ is the
source of the spurious singularities. Of course, once the full set of diagrams
is taken into account these singularities must cancel due to gauge invariance.
However at the intermediate steps they need regularisation.
The traditional approach \cite{Curci:1980uw,Hei98,Heinrich:1997kv,Ellis:1996nn}
is to use the principal value prescription:
\begin{equation}
\nonumber
\Bigl[\frac{1}{nl}\Bigr]_{PV} = \frac{nl}{(nl)^2 + \delta^2(np)^2}.
\end{equation}
As discussed in \cite{Curci:1980uw} this prescription is more like
a "phenomenological rule" than a theorem in QFT. 
A rigorous prescription has been proposed in
\cite{Mandelstam:1982cb,Leibbrandt:1983pj}. It uses
two auxiliary vectors, $n$ and $n^\star$. However this scheme is
difficult in practical calculations \cite{Bassetto:1998uv,Hei98,Heinrich:1997kv}
and, due to "ghosts", it is not suitable for stochastic applications%
\footnote{The concept of linear denominators, 
$1/(nl\pm i0)$, 
has been used also in NNLO
calculations of rapidity distributions of EW bosons \cite{Anastasiou:2003ds}
in the context of introducing the Dirac-delta constraint via formula:
\begin{equation}
\nonumber
\int d^mk\;\delta\Bigl(\frac{kp_1}{kp_2} -u\Bigr)
   \to
\int d^mk\;\frac{kp_2}{k(p_1-up_2) -i0} - c.c.
\end{equation} }.

\section{New use of PV prescription}
Separate contributions to the NLO non-singlet kernels calculated in the PV
prescription have $1/\epsilon^3$ singularities in dimensional
parameter $\epsilon=2-d/2$, see \cite{Curci:1980uw,Hei98,Heinrich:1997kv}. Only
once
the real-real and real-virtual graphs are
combined, these singularities cancel. This situation is not acceptable for the
Monte Carlo applications, which must be done in four dimensions.
Therefore in \cite{Jadach:2011kc}, we performed a calculation of the real-real
contributions in a modified PV scheme in which the $1/\epsilon^3$ poles
were replaced by $(1/\epsilon) \ln^2\delta$ terms. In order to fully define this
New PV (NPV) prescription we had to discuss the real-virtual corrections as well
\cite{Gituliar:2014eba}.
Based on these results we have formulated the new prescription as follows:
\begin{itemize}
\item
{\em Standard PV}: regularise {\em only the gluon propagator} with PV,
regularise all other singularities in (+)-components of integration momenta with
dimensional regularisation:
\begin{equation} 
\frac{d^ml}{l_+^{1-\epsilon}},\;\;\;\;\;\; l_+ = \frac{nl}{np}.
\nonumber
\end{equation}
\item
{\em NPV}: use PV to regularise {\em all singularities} of the
integrand in (+)-components of integration momenta, both real and virtual,
keeping higher order terms in $\epsilon$ as needed:
\begin{equation} 
\frac{d^ml}{l_+^{1-\epsilon}}\, \to d^ml\,
\biggl[\frac{1}{l_+}\biggr]_{PV} 
\Bigl(1+\epsilon\ln l_+ +\epsilon^2 \frac{1}{2}\ln^2l_+ +\dots\Bigr).
\end{equation}
\end{itemize}
The motivation of this approach is the following.
All (+)-singularities cancel in
the final expression (kernel) \cite{Ellis:1978ty,Ellis:1978sf}. Therefore it is
justified to extend the "phenomenological PV rule" of {\rm
Curci-Furmanski-Petronzio}, which is based on analogous cancellation of
"spurious" singularities in the final expression.

There is one significant consequence of the new scheme: now all the integrals,
including the non-axial (Feynman-type) ones depend on the axial vector
$n$ (via the products like $nq \equiv q^+$). Consider for example the
three-point scalar integral with kinematics $p^2=(p-q)^2=0$:
\begin{equation} \label{eq:def-feyn}
  J_3^\mathrm{F}
  =
  \int \frac{d^m l}{(2\pi)^m}
    \frac{1}{l^2 (q-l)^2 (p-l)^2}
\end{equation}
In the standard PV prescription we obtain:
\begin{equation}
  J_3^\mathrm{F}
  =
C
  \left(
    -\frac{1}{\epsilon^2} + \frac{\pi^2}{6}
  \right),\;\;\;\;\;\;\;\;
C = 
  i\frac{\Gamma(1-\epsilon)}{(4\pi)^2 \vert{q^2}\vert}
  \left(
    \frac{4\pi}{\vert{q^2}\vert}
  \right)^{-\epsilon}
\end{equation}
whereas in the NPV prescription the result is more complicated:
\begin{equation}
\label{eq:our-f}
  J_3^\mathrm{F}
  = 
  C\bigg(
   - \frac{1}{\epsilon}\Bigl( -2 \ln\delta + \ln(1-x)\Bigr)
    + 2 \ln^2\delta - 2 \ln\delta \ln(1-x) + \frac{1}{2}\ln^2(1-x)
  \bigg).
\end{equation}
Note, that the singularity $1/\epsilon^2$ has been replaced by
$(1/\epsilon)\ln\delta$ and $\ln^2\delta$.

\section{Exclusive contributions to $P_{qq}$ kernel}
\begin{figure}
  \centerline{
    \includegraphics[height=2.25cm]{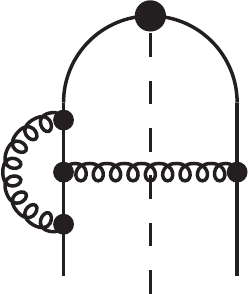}
    \hspace{1.5cm}
    \includegraphics[height=2.25cm]{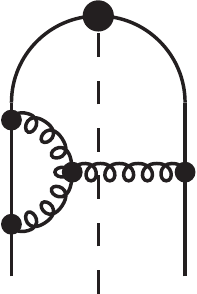}
    \hspace{1.3cm}
    \includegraphics[height=2.25cm]{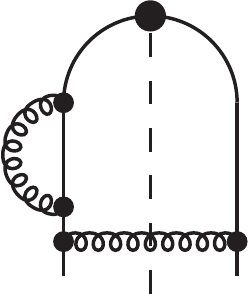}
    \hspace{1.5cm}
    \includegraphics[height=2.25cm]{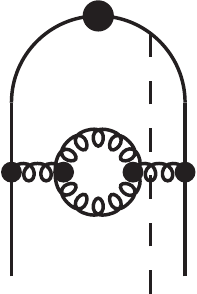}
    \hspace{1.5cm}
    \includegraphics[height=2.25cm]{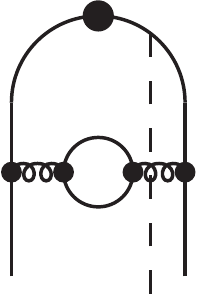}
  }
\centerline{      $(c):C_F^2-\frac{1}{2}C_FC_A$ \hskip 0.8cm 
                      $(d):\frac{1}{2}C_FC_A$ \hskip 1.6cm 
                      $(e):\frac{1}{2}C_F^2$  \hskip 1.6cm 
                      $(f):C_FC_A$ \hskip 1.4cm 
                      $(g):C_FT_F$ }
\caption{Real-virtual contributions to NLO non-singlet $P_{qq}$ kernel}
\label{fig1}
\end{figure}

In this section we will present the exclusive real-virtual
NLO contributions to the evolution kernel $P_{qq}$, obtained in the NPV
scheme.
Let us start with a brief reminder on how the kernels are calculated. The
main idea of collinear factorization is that the contributing graphs can be
grouped in a ladder-like structure of objects, $K$; which are
"two-particle-irreducible" (axial gauge is instrumental here). 
In the LO approximation the $K$ are just single emission graphs and the
whole structure is a genuine ladder. At the NLO level, among others the
real-virtual-type graphs contribute to $K$, see Fig.\ \ref{fig1}.
The (generalized) ladder is then cut into pieces by means of appropriate
projection operators,
which extract the singular parts of $K$. In the case of
fermions it is done simply by inserting the $\slashed{n}$ matrix and taking the
trace:
\begin{equation}
W_N = x\hbox{Tr}\Bigl[\frac{\slashed{n}}{4nq}\, K \, \slashed{p}\Bigr].
\end{equation}
The $\slashed{n}$ matrices are represented by the black dots on top of lines in
Fig.\ \ref{fig1}. The $\slashed{p}$ matrices closing fermion lines from below
are not shown.
In this way we obtain the non-integrated partonic density $W_N$. In the next
step the
integral over virtual loop momenta hidden in $K$ is performed and the
renormalisation procedure is carried out. Finally we end up with the exclusive
(renormalized) partonic density $W_R$, which depends on the four-momentum $k$ of
the real emitted parton. Once the real phase space $d\Phi(k)$ is
integrated out we obtain the standard inclusive parton densities, $\tilde
\Gamma$.
The evolution kernel $P_{qq}$ is then related to $\tilde \Gamma$ as
follows (up to terms proportional to $\delta(1-x)$):
\begin{equation}
\tilde \Gamma_{qq}(x,\epsilon)
 = \delta(1-x)
  + \frac{1}{\epsilon}\biggl(
            \frac{\alpha_S}{2\pi} P_{qq}^{LO}(x)
       +\frac{1}{2}\Bigl( \frac{\alpha_S}{2\pi} \Bigr)^2  P_{qq}^{NLO}(x)
      +\dots \biggr)
  + {\cal O}\Bigl(\frac{1}{\epsilon^2}\Bigr).
\end{equation}
Let us now present the complete exclusive partonic densities $W_R$ for the
non-singlet $qq$ case in the NPV scheme corresponding to Fig.\
\ref{fig1}. We group them in two color structures: $C_F^2$ and $C_FC_A$ plus
$C_FT_F$:
\begin{eqnarray}
W_R^{C_F^2}=
  \alpha_S^2 \: C_F^2 \: \frac{\Gamma(1-\epsilon)}{(4\pi)^\epsilon} \:
\frac{1}{{|q^2|}} \;
  \biggl\{ &&
    \frac{1}{\epsilon} \: 4\ln{x} \left(
\left(\frac{{|q^2|}}{\mu_R^2}\right)^\epsilon - 1 \right) P_{qq} 
\nonumber
\\
   &&  + \Big( p_{qq} \; 4 \: \Li(1-x)  - (1-x) + (1+x) \Big)
\left(\frac{{|q^2|}}{\mu_R^2}\right)^\epsilon
  \biggr\},
\end{eqnarray}
\begin{eqnarray}
W_R^{C_FC_A+C_FT_F}=
  && 
  \alpha_S^2 \: C_F \: \frac{\Gamma(1-\epsilon)}{(4\pi)^\epsilon} \:
\frac{1}{{|q^2|}} \:
  \biggl\{
    \frac{1}{\epsilon} \left[ C_A \frac{11}{3} - T_F \frac{4}{3} - 4 \: C_A \:
\big(\ln(1-x) + I_0 \big) \left(\frac{{|q^2|}}{\mu_R^2}\right)^\epsilon 
                      \right] P_{qq} \nonumber 
\\
    && \qquad + 4 \: C_A \biggl[ p_{qq} \Big(\Li(1) - \Li(1-x) + I_0 \ln(1-x) 
  -2 I_1 \Big) - \frac{x}{2} \biggr] 
  \left(\frac{{|q^2|}}{\mu_R^2}\right)^\epsilon
  \biggr\},
\label{eq:Gamma1_CACF_excl}
\\ 
P_{qq}= &&p_{qq} + \epsilon(1-x),\quad p_{qq}= \frac{1+x^2}{1-x},
\quad
I_0 = -\ln\delta,\quad I_1 = -\frac{1}{2}\ln^2\delta - \frac{\pi^2}{24}
\end{eqnarray}
where $q=p-k$ is the momentum of the virtual quark after real emission. As one
can
see the double poles in $\epsilon$ vanished, or more precisely, have been
replaced by the $\ln\delta$-type terms. 
This is the main difference with
respect to the results in the standard PV scheme known from the literature, and
it makes these results "Monte Carlo friendly".

The exclusive densities $W_R^{C_F^2}$ and $W_R^{C_FC_A+C_FT_F}$ can be further
integrated over the remaining one-particle phase space. In this way one obtains
the
inclusive densities, and hence the standard NLO evolution kernel $P_{qq}$. We
have
verified that in the NPV scheme we reproduce the $\overline{\text{MS}}$
kernel $P_{qq}$ from the literature. Moreover, the
inclusive NPV results agree with the PV ones also on a graph-by-graph basis,
provided the real-real and real-virtual components are added. This forms a
strong cross-check of our scheme. Complete results in the NPV scheme, for each
graph separately and both exclusive and inclusive, can be found in
\cite{Gituliar:2014mua}.

\section{ The {\tt Axiloop} package}
In order to facilitate symbolic calculations in the axial (light-cone) gauge we
have written a Mathematica package named {\tt Axiloop}. 
It is
primarily oriented around automated calculation of the NLO kernels. The code is
publicly available at
\url{http://gituliar.org/axiloop/}

Let us list the main features of this package:

$\bullet$ Contains a library of integrals (scalar, vector and tensor), in PV and
NPV prescriptions.

$\bullet$ Performs one-loop integration and renormalisation (keeping track of
the UV and IR poles).

$\bullet$ Performs one-particle final-state integration.

$\bullet$ Provides auxiliary routines, e.g.\ for Passarino-Veltman reduction.
\\
As an output one can obtain all mentioned earlier types of density functions
(not integrated, bare, exclusive renormalized, inclusive as well as
counter-terms).
\section{Summary}
We presented the status of the {\tt KrkMC} project of constructing QCD Parton
Shower
in which both the hard matrix element and the shower are upgraded to NLO
precision
level. We briefly outlined the strategy of the upgrade, and then we discussed in
detail one of the components; the evolution kernels that had to be
recalculated in a new way.

We presented calculation of all NLO real-virtual one-loop
components of the non-singlet $P_{qq}$ kernel in a new, MC-friendly,
regularisation
scheme, both in inclusive and exclusive forms.
The new scheme is based on modified usage of the PV prescription in
the light-cone
gauge: it applies PV regularisation to all the singularities in the plus
component of the integration momenta. 

In the NPV prescription the inclusive NLO kernel $P_{qq}$ agrees with the
standard PV one. However, separate real-real and real-virtual contributions
differ in PV and NPV: the $1/\epsilon^3$ poles, present in
PV, are replaced by $(1/\epsilon)\ln^2\delta$ etc. 
As a consequence, in the NPV scheme no cancellation of $1/\epsilon^3$ terms
between real and virtual components occurs. Moreover, most of the real
graphs are free of higher order
poles and due to that can be calculated in
four dimensions and can be used for the stochastic simulations.

\section*{Acknowledgments}
This work has been partly supported by 
 the Polish National Science Center grants DEC-2011/03/B/ST2/02632
and UMO-2012/04/M/ST2/00240,
  the Research Executive Agency (REA) of the European Union 
  Grant PITN-GA-2010-264564 (LHCPhenoNet),
the U.S.\ Department of Energy
under grant DE-FG02-13ER41996 and the Lightner-Sams Foundation.



\providecommand{\href}[2]{#2}\begingroup\raggedright\endgroup

\end{document}